\input tables
\newif\iflanl
\openin 1 lanlmac
\ifeof 1 \lanlfalse \else \lanltrue \fi
\closein 1
\iflanl
    \input lanlmac
\else
    \message{[lanlmac not found - use harvmac instead}
    \input harvmac
\fi
\newif\ifhypertex
\ifx\hyperdef\UnDeFiNeD
    \hypertexfalse
    \message{[HYPERTEX MODE OFF}
    \def\hyperref#1#2#3#4{#4}
    \def\hyperdef#1#2#3#4{#4}
    \def\hypernoname{}
    \def\e@tf@ur#1{}
    \def\eprt#1{{\tt #1}}
    \def\CERN{\address{CERN, CH--1211 Geneva 23, Switzerland}}
    \def\wl{W.\ Lerche}
\else
    \hypertextrue
    \message{[HYPERTEX MODE ON}
\def\eprt#1{{\tt
#1}}
\def\CERN{\address{

Theory Division, CERN, Geneva, Switzerland}}
\def\wl{
 W.\ Lerche}
\fi
\newif\ifdraft

\noblackbox
\catcode`\@=11
\newif\iffrontpage
\ifx\answ\bigans
\def\titleft{\titla}
\magnification=1200\baselineskip=14pt plus 2pt minus 1pt
%
\advance\hoffset by-0.075truein
\advance\voffset by1.truecm
\hsize=6.15truein\vsize=600.truept\hsbody=\hsize\hstitle=\hsize
\else\let\lr=L
\def\titleft{\titla}
\magnification=1000\baselineskip=14pt plus 2pt minus 1pt
%
\hoffset=-0.75truein\voffset=-.0truein
\vsize=6.5truein
\hstitle=8.truein\hsbody=4.75truein
\fullhsize=10truein\hsize=\hsbody
\fi
\parskip=4pt plus 15pt minus 1pt
%
\newif\iffigureexists
\newif\ifepsfloaded
\def\epsfcheck{
\ifdraft
\input epsf\epsfloadedtrue
\else
  \openin 1 epsf
  \ifeof 1 \epsfloadedfalse \else \epsfloadedtrue \fi
  \closein 1
  \ifepsfloaded
    \input epsf
  \else
\immediate\write20{NO EPSF FILE --- FIGURES WILL BE IGNORED}
  \fi
\fi
\def\epsfcheck{}}
\def\checkex#1{
\ifdraft
\figureexistsfalse\immediate%
\write20{Draftmode: figure #1 not included}
\figureexiststrue
\else\relax
    \ifepsfloaded \openin 1 #1
        \ifeof 1
           \figureexistsfalse
  \immediate\write20{FIGURE FILE #1 NOT FOUND}
        \else \figureexiststrue
        \fi \closein 1
    \else \figureexistsfalse
    \fi
\fi}
\def\missbox#1#2{$\vcenter{\hrule
\hbox{\vrule height#1\kern1.truein
\raise.5truein\hbox{#2} \kern1.truein \vrule} \hrule}$}
\def\lfig#1{
\let\labelflag=#1%
\def\numb@rone{#1}%
\ifx\labelflag\UnDeFiNeD%
{\xdef#1{\the\figno}%
\writedef{#1\leftbracket{\the\figno}}%
\global\advance\figno by1%
}\fi{\hyperref{}{figure}{{\numb@rone}}{Fig.{\numb@rone}}}}
\def\figinsert#1#2#3#4{
\epsfcheck\checkex{#4}%
\def\figsize{#3}%
\let\flag=#1\ifx\flag\UnDeFiNeD
{\xdef#1{\the\figno}%
\writedef{#1\leftbracket{\the\figno}}%
\global\advance\figno by1%
}\fi
\goodbreak\midinsert%
\iffigureexists
\centerline{\epsfysize\figsize\epsfbox{#4}}%
\else%
\vskip.05truein
  \ifepsfloaded
  \ifdraft
  \centerline{\missbox\figsize{Draftmode: #4 not included}}%
  \else
  \centerline{\missbox\figsize{#4 not found}}
  \fi
  \else
  \centerline{\missbox\figsize{epsf.tex not found}}
  \fi
\vskip.05truein
\fi%
{\smallskip%
\leftskip 4pc \rightskip 4pc%
\noindent\ninepoint\sl \baselineskip=11pt%
{\bf{\hyperdef\hypernoname{figure}{{#1}}{Fig.{#1}}}:~}#2%
\smallskip}\bigskip\endinsert%
}

\def\boxit#1{\vbox{\hrule\hbox{\vrule\kern8pt
\vbox{\hbox{\kern8pt}\hbox{\vbox{#1}}\hbox{\kern8pt}}
\kern8pt\vrule}\hrule}}
\def\mathboxit#1{\vbox{\hrule\hbox{\vrule\kern8pt\vbox{\kern8pt
\hbox{$\displaystyle #1$}\kern8pt}\kern8pt\vrule}\hrule}}
%
\font\bigit=cmti10 scaled \magstep1

\font\titla=cmr10 scaled\magstep3
\font\tenmss=cmss10
\font\absmss=cmss10 scaled\magstep1

\newfam\mssfam
\font\footrm=cmr8  \font\footrms=cmr5
\font\footrmss=cmr5   \font\footi=cmmi8
\font\footis=cmmi5   \font\footiss=cmmi5
\font\footsy=cmsy8   \font\footsys=cmsy5
\font\footsyss=cmsy5   \font\footbf=cmbx8
\font\footmss=cmss8
\def\footfont{\def\rm{\fam0\footrm}
\textfont0=\footrm \scriptfont0=\footrms
\scriptscriptfont0=\footrmss
\textfont1=\footi \scriptfont1=\footis
\scriptscriptfont1=\footiss
\textfont2=\footsy \scriptfont2=\footsys
\scriptscriptfont2=\footsyss
\textfont\itfam=\footi \def\it{\fam\itfam\footi}
\textfont\mssfam=\footmss \def\mss{\fam\mssfam\footmss}
\textfont\bffam=\footbf \def\bf{\fam\bffam\footbf} \rm}
\def\tenpoint{\def\rm{\fam0\tenrm}
\textfont0=\tenrm \scriptfont0=\sevenrm
\scriptscriptfont0=\fiverm
\textfont1=\teni  \scriptfont1=\seveni
\scriptscriptfont1=\fivei
\textfont2=\tensy \scriptfont2=\sevensy
\scriptscriptfont2=\fivesy
\textfont\itfam=\tenit \def\it{\fam\itfam\tenit}
\textfont\mssfam=\tenmss \def\mss{\fam\mssfam\tenmss}
\textfont\bffam=\tenbf \def\bf{\fam\bffam\tenbf} \rm}
\ifx\answ\bigans\def\abstractfont{\tenpoint}\else
\def\abstractfont{\def\rm{\fam0\absrm}
\textfont0=\absrm \scriptfont0=\absrms
\scriptscriptfont0=\absrmss
\textfont1=\absi \scriptfont1=\absis
\scriptscriptfont1=\absiss
\textfont2=\abssy \scriptfont2=\abssys
\scriptscriptfont2=\abssyss
\textfont\itfam=\bigit \def\it{\fam\itfam\bigit}
\textfont\mssfam=\absmss \def\mss{\fam\mssfam\absmss}
\textfont\bffam=\absbf \def\bf{\fam\bffam\absbf}\rm}\fi
%
\def\f@@t{\baselineskip10pt\lineskip0pt\lineskiplimit0pt
\bgroup\aftergroup\@foot\let\next}
\setbox\strutbox=\hbox{\vrule height 8.pt depth 3.5pt width\z@}
\def\vfootnote#1{\insert\footins\bgroup
\baselineskip10pt\footfont
\interlinepenalty=\interfootnotelinepenalty
\floatingpenalty=20000
\splittopskip=\ht\strutbox \boxmaxdepth=\dp\strutbox
\leftskip=24pt \rightskip=\z@skip
\parindent=12pt \parfillskip=0pt plus 1fil
\spaceskip=\z@skip \xspaceskip=\z@skip
\Textindent{$#1$}\footstrut\futurelet\next\fo@t}
\def\Textindent#1{\noindent\llap{#1\enspace}\ignorespaces}
\def\foot{\global\advance\ftno by1%
\attach{\hyperref{}{footnote}{\the\ftno}{\footsymbolgen}}%
\vfootnote{\hyperdef\hypernoname{footnote}{\the\ftno}{\footsymbol}}}%
\def\footnote#1{\global\advance\ftno by1%
\attach{\hyperref{}{footnote}{\the\ftno}{#1}}%
\vfootnote{\hyperdef\hypernoname{footnote}{\the\ftno}{#1}}}%
\newcount\lastf@@t           \lastf@@t=-1
\newcount\footsymbolcount    \footsymbolcount=0
\global\newcount\ftno \global\ftno=0
\def\footsymbolgen{\relax\footsym
\global\lastf@@t=\pageno\footsymbol}
\def\footsym{\ifnum\footsymbolcount<0
\global\footsymbolcount=0\fi
{\iffrontpage \else \advance\lastf@@t by 1 \fi
\ifnum\lastf@@t<\pageno \global\footsymbolcount=0
\else \global\advance\footsymbolcount by 1 \fi }
\ifcase\footsymbolcount
\fd@f\dagger\or \fd@f\diamond\or \fd@f\ddagger\or
\fd@f\natural\or \fd@f\ast\or \fd@f\bullet\or
\fd@f\star\or \fd@f\nabla\else \fd@f\dagger
\global\footsymbolcount=0 \fi }
\def\fd@f#1{\xdef\footsymbol{#1}}
\def\space@ver#1{\let\@sf=\empty \ifmmode #1\else \ifhmode
\edef\@sf{\spacefactor=\the\spacefactor}
\unskip${}#1$\relax\fi\fi}
\def\attach#1{\space@ver{\strut^{\mkern 2mu #1}}\@sf}
%
\newif\ifnref
\def\rrr#1#2{\relax\ifnref\nref#1{#2}\else\ref#1{#2}\fi}
\def\ldf#1#2{\begingroup\obeylines
\gdef#1{\rrr{#1}{#2}}\endgroup\unskip}
\def\nrf#1{\nreftrue{#1}\nreffalse}
\def\doubref#1#2{\refs{{#1},{#2}}}
\def\multref#1#2#3{\nrf{#1#2#3}\refs{#1{--}#3}}
\nreffalse
\def\refout{\listrefs}

\def\lref{\ldf}

\def\eqn#1{\xdef #1{(\noexpand\hyperref{}%
{equation}{\secsym\the\meqno}%
{\secsym\the\meqno})}\eqno(\hyperdef\hypernoname{equation}%
{\secsym\the\meqno}{\secsym\the\meqno})\eqlabeL#1%
\writedef{#1\leftbracket#1}\global\advance\meqno by1}
\def\eqnalign#1{\xdef #1{\noexpand\hyperref{}{equation}%
{\secsym\the\meqno}{(\secsym\the\meqno)}}%
\writedef{#1\leftbracket#1}%
\hyperdef\hypernoname{equation}%
{\secsym\the\meqno}{\e@tf@ur#1}\eqlabeL{#1}%
\global\advance\meqno by1}
\def\eqnalign#1{\xdef #1{(\secsym\the\meqno)}
\writedef{#1\leftbracket#1}%
\global\advance\meqno by1 #1\eqlabeL{#1}}
%

%
\def\chap#1{\newsec{#1}}
\def\chapter#1{\chap{#1}}
\def\sect#1{\subsec{#1}}
\def\section#1{\sect{#1}}
\def\\{\ifnum\lastpenalty=-10000\relax
\else\hfil\penalty-10000\fi\ignorespaces}
\def\note#1{\leavevmode%
\edef\@@marginsf{\spacefactor=\the\spacefactor\relax}%
\ifdraft\strut\vadjust{%
\hbox to0pt{\hskip\hsize%
\ifx\answ\bigans\hskip.1in\else\hskip .1in\fi%
\vbox to0pt{\vskip-\dp
\strutbox\sevenbf\baselineskip=8pt plus 1pt minus 1pt%
\ifx\answ\bigans\hsize=.7in\else\hsize=.35in\fi%
\tolerance=5000 \hbadness=5000%
\leftskip=0pt \rightskip=0pt \everypar={}%
\raggedright\parskip=0pt \parindent=0pt%
\vskip-\ht\strutbox\noindent\strut#1\par%
\vss}\hss}}\fi\@@marginsf\kern-.01cm}
\def\titlepage{%
\frontpagetrue\nopagenumbers\abstractfont%
\hsize=\hstitle\rightline{\vbox{\baselineskip=10pt%
{\abstractfont\pubnum}}}\pageno=0}
\frontpagefalse
\def\pubnum{}
\def\pdate{\number\month/\number\yearltd}
\def\makefootline{\iffrontpage\vskip .27truein
\line{\the\footline}
\vskip -.1truein\leftline{\vbox{\baselineskip=10pt%
{\abstractfont\pdate}}}
\else\vskip.5cm\line{\hss \tenrm $-$ \folio\ $-$ \hss}\fi}
\def\title#1{\vskip .7truecm\titlestyle{\titleft #1}}
\def\titlestyle#1{\par\begingroup \interlinepenalty=9999
\leftskip=0.02\hsize plus 0.23\hsize minus 0.02\hsize
\rightskip=\leftskip \parfillskip=0pt
\hyphenpenalty=9000 \exhyphenpenalty=9000
\tolerance=9999 \pretolerance=9000
\spaceskip=0.333em \xspaceskip=0.5em
\noindent #1\par\endgroup }
\def\autskip{\ifx\answ\bigans\vskip.5truecm\else\vskip.1cm\fi}
\def\author#1{\vskip .7in \centerline{#1}}

\def\address#1{\ifx\answ\bigans\vskip.2truecm
\else\vskip.1cm\fi{\it \centerline{#1}}}
\def\abstract#1{
\vskip .5in\vfil\centerline
{\bf Abstract}\penalty1000
{{\smallskip\ifx\answ\bigans\leftskip 2pc \rightskip 2pc
\else\leftskip 5pc \rightskip 5pc\fi
\noindent\abstractfont \baselineskip=12pt
{#1} \smallskip}}
\penalty-1000}
\def\endpage{\tenpoint\supereject\global\hsize=\hsbody%
\frontpagefalse\footline={\hss\tenrm\folio\hss}}
\def\ack{\vskip2.cm\centerline{{\bf Acknowledgements}}}
%
%

%
\def\bfone{\relax{\rm 1\kern-.35em 1}}
\def\inbar{\vrule height1.5ex width.4pt depth0pt}
\def\IC{\relax\,\hbox{$\inbar\kern-.3em{\mss C}$}}
\def\ID{\relax{\rm I\kern-.18em D}}
\def\IF{\relax{\rm I\kern-.18em F}}
\def\IH{\relax{\rm I\kern-.18em H}}
\def\II{\relax{\rm I\kern-.17em I}}
\def\IN{\relax{\rm I\kern-.18em N}}
\def\IP{\relax{\rm I\kern-.18em P}}
\def\IQ{\relax\,\hbox{$\inbar\kern-.3em{\rm Q}$}}
\def\IR{\relax{\rm I\kern-.18em R}}
\font\cmss=cmss10 \font\cmsss=cmss10 at 7pt
\def\ZZ{\relax\ifmmode\mathchoice
{\hbox{\cmss Z\kern-.4em Z}}{\hbox{\cmss Z\kern-.4em Z}}
{\lower.9pt\hbox{\cmsss Z\kern-.4em Z}}
{\lower1.2pt\hbox{\cmsss Z\kern-.4em Z}}\else{\cmss Z\kern-.4em
Z}\fi}
  \def\d{\delta}

\def\cF{{\cal F}}

 \def\cM{{\cal M}}

\def\nup#1({Nucl.\ Phys.\ $\us {B#1}$\ (}
\def\plt#1({Phys.\ Lett.\ $\us  {#1}$\ (}
\def\cmp#1({Comm.\ Math.\ Phys.\ $\us  {#1}$\ (}
\def\prp#1({Phys.\ Rep.\ $\us  {#1}$\ (}
\def\prl#1({Phys.\ Rev.\ Lett.\ $\us  {#1}$\ (}
\def\prv#1({Phys.\ Rev.\ $\us  {#1}$\ (}
\def\mpl#1({Mod.\ Phys.\ Let.\ $\us  {A#1}$\ (}
\def\ijmp#1({Int.\ J.\ Mod.\ Phys.\ $\us{A#1}$\ (}
\def\tit#1|{{\it #1},\ }
%

%

\def\ni{\noindent}
\def\tilde{\widetilde}

\def\us#1{\underline{#1}}

\def\hat{\widehat}

\def\Coeff#1#2{{#1\over #2}}
\def\Coe#1.#2.{{#1\over #2}}

\def\coe#1.#2.{\relax{\textstyle {#1 \over #2}}\displaystyle}

\def\shalf{\relax{\textstyle {1 \over 2}}\displaystyle}

\def\to{\rightarrow}
\def\notin{\hbox{{$\in$}\kern-.51em\hbox{/}}}

\def\del{\partial}


\catcode`\@=12


\def\tr {{\rm Tr}}


 1


\def\frac#1#2{{#1\over #2}}
\def\n#1{{\bf q}^{(#1)}_\infty}

\def\xone{\IP(1,1,2,2,6)[12]}
\def\xtwo{\IP(1,1,2,2,2)[8]}
\def\Xone{$\xone$}
\def\Xtwo{$\xtwo$}

\def\Inters{I}
\def\Interg{\Inters^{{\rm geom}}}

\def\to{{t_1}}
\def\tz{{t_2}}

\def\nn#1{{n_{#1}}}
\def\n#1#2{n_{#1}^{#2}}

\def\chv#1{{\rm ch}_{#1}(V)}
\def\IG{\relax\,\hbox{$\inbar\kern-.3em{\mss G}$}}

    \def\pk{P.\ Kaste}
    \def\cl{C.A.\ L\"utken}
    \def\jw{J.\ Walcher}


\def\nihil#1{{\sl #1}}
\def\br{\hfill\break}

\def\ijmp {{Int. J. Mod. Phys.\ }{\bf A}}

\lref\COFKM{P.\ Candelas, X.\ De La Ossa, A.\ Font, 
S.\ Katz and D.R.\ Morrison, 
\nihil{Mirror symmetry for two parameter models.\ I,}
 Nucl.\ Phys.\ {\bf B416} 481 (1994), 
\eprt{hep-th/9308083}. 
}

\lref\KKLMV{S.\ Kachru, A.\ Klemm, W.\ Lerche, P.\ Mayr and C.\ Vafa, 
\nihil{Nonperturbative results on the point particle 
limit of $N=2$ heterotic string compactifications,}
 Nucl.\ Phys.\ {\bf B459} 537 (1996), 
\eprt{hep-th/9508155}. 
}

\lref\DDCR{D.\ Diaconescu and C.\ R\"omelsberger, 
\nihil{D-branes and bundles on elliptic fibrations,}
\eprt{hep-th/9910172}. 
}

\lref\HM{J.A.\ Harvey and G.\ Moore, 
\nihil{On the algebras of BPS states,}
 Commun.\ Math.\ Phys.\ {\bf 197} 489 (1998), 
\eprt{hep-th/9609017}. 
}

\lref\mukai{S. Mukai, \nihil{Symplectic structure of the moduli of sheaves
on an abelian or K3 surface,} Invent. Math. {\bf 77}(1984) 101; 
\nihil{On the moduli space of bundles on K3 surfaces, I} in {\it Vector
Bundles on Algebraic Varieties,} Tata Inst.\ of Fund.\ Research.}

\lref\BDLR{I.\ Brunner, M.R.\ Douglas, A.\ Lawrence and 
C.\ R\"omelsberger, 
\nihil{D-branes on the quintic,}
\eprt{hep-th/9906200}. 
}

\lref\kthreefib{
{A.\ Klemm, W.\ Lerche and P.\ Mayr, 
\nihil{K3 Fibrations and heterotic type II string duality,}
 Phys.\ Lett.\ {\bf B357} 313 (1995), 
\eprt{hep-th/9506112}; 
}
\br
{P.\ S.\ Aspinwall and J.\ Louis, 
\nihil{On the Ubiquity of K3 Fibrations in String Duality,}
 Phys.\ Lett.\ {\bf B369} 233 (1996), 
\eprt{hep-th/9510234}. 
}
}

\lref\GS{
{M.\ Gutperle and Y.\ Satoh, 
\nihil{D-branes in Gepner models and supersymmetry,}
 Nucl.\ Phys.\ {\bf B543} 73 (1999), 
\eprt{hep-th/9808080};
}
{
\nihil{D0-branes in Gepner models and N = 2 black holes,}
 Nucl.\ Phys.\ {\bf B555} 477 (1999), 
\eprt{hep-th/9902120}. 
}
}

\lref\RS{A.\ Recknagel and V.\ Schomerus, 
\nihil{D-branes in Gepner models,}
 Nucl.\ Phys.\ {\bf B531} 185 (1998), 
\eprt{hep-th/9712186}. 
}

\lref\RSb{
{A.\ Recknagel and V.\ Schomerus, 
\nihil{Boundary deformation theory and moduli spaces of D-branes,}
 Nucl.\ Phys.\ {\bf B545} 233 (1999), 
\eprt{hep-th/9811237};
}
{ 
\nihil{Moduli spaces of D-branes in CFT-backgrounds,}
\eprt{hep-th/9903139}. 
}
}

\lref\MD{M.R.\ Douglas, 
\nihil{Topics in D-geometry,}
\eprt{hep-th/9910170}. 
}

\lref\cumrun{C.\ Vafa, 
\nihil{Extending mirror conjecture to Calabi-Yau with bundles,}
\eprt{hep-th/9804131}. 
}

\lref\OOY{H.\ Ooguri, Y.\ Oz and Z.\ Yin, 
\nihil{D-branes on Calabi-Yau spaces and their mirrors,}
 Nucl.\ Phys.\ {\bf B477} 407 (1996), 
\eprt{hep-th/9606112}. 
}

\lref\GHM{M.\ B.\ Green, J.\ A.\ Harvey and G.\ Moore, 
\nihil{I-brane inflow and anomalous couplings on D-branes,}
 Class.\ Quant.\ Grav.\ {\bf 14} 47 (1997), 
\eprt{hep-th/9605033}. 
}

\lref\DDJG{D.\ Diaconescu and J.\ Gomis, 
\nihil{Fractional branes and boundary states in orbifold theories,}
\eprt{hep-th/9906242}. 
}

\lref\BS{M.\ Bershadsky and V.\ Sadov, 
\nihil{F-theory on K3 x K3 and instantons on 7-branes,}
 Nucl.\ Phys.\ {\bf B510} 232 (1998), 
\eprt{hep-th/9703194}. 
}

\lref\GJS{S.\ Govindarajan, T.\ Jayaraman and T.\ Sarkar, 
\nihil{World sheet approaches to D-branes on supersymmetric cycles,}
\eprt{hep-th/9907131}. 
}

\lref\FS{J.\ Fuchs and C.\ Schweigert, 
\nihil{Branes: From free fields to general backgrounds,}
 Nucl.\ Phys.\ {\bf B530} 99 (1998), 
\eprt{hep-th/9712257}. 
}



\def\pubnum{
\hbox{CERN-TH/99-398}
\hbox{hep-th/9912147}
\hbox{}}
\def\pdate{}
\titlepage
\vskip2.cm
\title
{{\titlefont D-Branes on K3-Fibrations}}
\vskip -.7cm
\autskip
\author{\pk, \wl,\ \cl\footnote{1\ }{On leave from Dept. of Physics,
University of Oslo, N-0316 Oslo, Norway}\ and 
\jw\footnote{2\ }{Also at Institut f\"ur Theoretische Physik, 
ETH-H\"onggerberg, CH-8093 Z\"urich, Switzerland}}
\vskip0.2truecm
\CERN
\vskip-.4truecm

\abstract{
$B$-type $D$-branes are constructed on two different $K3$-fibrations
over $\IP_1$ using boundary conformal field theory at the rational
Gepner points of these models. The microscopic CFT charges  are
compared with the Ramond charges of $D$-branes wrapped on holomorphic
cycles of the corresponding  Calabi-Yau  manifold.  We study in
particular $D4$-branes and bundles localized on the $K3$ fibers, and
find from CFT that each irreducible component of a bundle on $K3$ gains
one modulus upon fibration over $\IP_1$. This is in agreement with
expectations and so provides a further test of the boundary CFT
approach to $D$-brane physics.
}

\vfil
\vskip 1.cm
\ni {CERN-TH/99-398}\hfill\break
\ni December 1999
\endpage
\baselineskip=14pt plus 2pt minus 1pt

\sequentialequations

\chapter{Introduction}

In the `large radius' region of moduli space, strings can be viewed
as moving in a fixed extrinsic background space-time whose geometry is
determined by the closed string sector of the theory itself; in
particular it must be a Calabi-Yau (CY) space. At string tree level
these classical geometric data and some of the  quantum corrections are
encoded in $N=2$ superconformal field  theory on the plane, hereafter
called `bulk' CFT.   At special points in its moduli space the CFT is
rational  and can therefore be exhaustively studied using Gepner's
construction. The complementary viewpoints provided by the `intrinsic,
microscopic' CFT picture, and the `extrinsic, macroscopic' CY
sigma-model, have provided a wealth of information about the closed
string sector, most notably the phenomenon of mirror symmetry,
amounting to a profound redirection of our thinking about the
structure of space-time at small scales.

Our understanding of the open string sector is currently undergoing a
parallel evolution which may ultimately provide a similar redirection
of our thinking about the gauge structure of the theory. Indeed, a
major effort is currently underway to  extend the mirror map to the
gauge bundles that may decorate various cycles in CY manifolds 
(see e.g., \doubref\OOY\cumrun).

For the open string sector the appropriate microscopic description is
provided by boundary conformal field theory on the upper half plane
(BCFT), and the corresponding macroscopic  objects are $D$-branes
wrapped on cycles of the CY target spaces embodied in the closed string
sector. Since open strings can end on $D$-branes, these should
be intimately related to the boundary states of the (microscopic) BCFT.
In fact they are supposedly nothing but manifestations of  the boundary
states at large radii of the compactification manifold, where naive
geometrical concepts apply. On the other hand, at small radii instanton
corrections can swamp out any classical picture, to an extent where the
very notion of  space-time, as well as the vector bundles it supports,
evaporates.

However, it is conceivable that the microscopic BCFT still encodes some
of the geometric information pertaining to the $D$-branes at large
radii, like the intersection properties of the cycles around which they
wrap.  This can be tested at the ``Gepner'' points in moduli space
where the theory is exactly solvable, and thus opens a window on
$D$-brane physics in a regime where quantum corrections are strong.
This approach has been pioneered in refs.\
\multref\BDLR{\MD\DDJG}\DDCR, building on earlier work, for example
\multref\RS{\FS\GS}\GJS.

Our purpose in this letter is to apply the methods developed in these
papers to study properties of $D$-branes on $K3$-fibrations. The
motivation is twofold. First, $K3$-fibrations are important in the
context of heterotic-type II duality \kthreefib, and their quantum
algebraic geometry  is well understood.  In particular, for some cases
the periods, monodromy and analytic continuation to Gepner points are
either directly available from the literature \COFKM, or can be
obtained with a reasonable amount of effort. Second, for generic
threefolds the quantum corrections at the Gepner point are strong, and
for some states \refs{\BDLR{-}\DDCR}\ there is no good match between
BCTF data at the Gepner point and geometrical $D$-brane data at large
radii -- for these cases it may not always be clear whether the quantum
geometries really differ so much, or whether the method fails.
However, for $K3$ fibrations we may expect that some quantities are
protected by non-renormalization properties of the $K3$, while the
whole theory is not as trivial as if we would be looking just at an
isolated $K3$. Therefore, studying $K3$-fibrations might not only offer
new insights in $D$-brane physics at small radii, but also provide
further tests of the validity of the BCFT method.

In Sect.\ 2, we review the BCFT approach at the Gepner point of
generic $n$-folds, but also improve and clarify some steps in the papers
\refs{\BDLR{-}\DDCR}. Where our work overlaps, we are in
agreement, and we explain why this is so.  In Sect.\ 3 we then discuss
some basic $D$-geometric features of the two $K3$-fibrations we study.
In section four, we compare the BCFT and the geometrical data,
producing lists of $D$-brane charges that correspond to the CFT
boundary states. Finally, in Sect.\ 4 we discuss some features of
these data and find agreement with expectations.

\chapter{Gepner models and boundary conformal field theory}

The Gepner construction is a GSO-projected    tensor product of $r$
minimal models at levels $k_i$,  whose central charges
$c_i=3k_i/(k_i+2)$ add up to  $3n\ (n = 1,2,3,\dots)$, suitably twisted
to enforce modular invariance of the partition function.  An explicit
connection with the  geometric data at `large radius' is provided by
the Landau-Ginzburg  representation of minimal models, provided that
the tensor product is extended by a trivial factor $k_{r+1} = 0$ when
$n + r$ is odd. Let $K = {\rm lcm}\{h_1,\dots ,h_r\}$, where   $h_i =
k_i + 2$ are called the heights of the minimal models. The $w_i =
K/h_i$ will here be called the (relative) weights of  the minimal
models in the given product.   Since $K$ is even when $n + r$ is odd,
$h_{r+1} = 2$ is automatically a  divisor of $K$ and the weight of the
trivial extension is $w_{r+1} = K/2$.   The smooth CY $n$-fold is then
obtained by desingularizing the degree $K$  quasi-homogeneous
hypersurface in the weighted projective space of  dimension ${\rm
dim}(w) - 1$ with scaling weights $w_i$.

We consider here only the two Gepner models 
$10^2 4^2 0/\ZZ_{12} = 10\otimes 10\otimes 4\otimes 4\otimes 0/\ZZ_{12}$ 
and $6^2 2^3/\ZZ_8$.  By the above algorithm, the
first model contains the CY manifold which is defined classically 
by resolving the singularities of a quasi-homogeneous polynomial 
hypersurface of total degree 12 in a four-dimensional projective 
space weighted by $(1,1,2,2,6)$, hereafter denoted \Xone\ for short.  
The second one contains the geometric data for a degree 8 projective variety,
denoted \Xtwo\ for short.  
Both these varieties are $K3$-fibrations over $\IP_1$.
Their quantum geometry has been extensively investigated using mirror 
symmetry \doubref\COFKM\KKLMV, and we will draw heavily on this information.

Open string states satisfying Neumann or Dirichlet 
boundary conditions are built from  generalized coherent states
(Ishibashi states) of a boundary conformal field theory.
The coherent basis $\vert i\rangle\rangle_{\Omega} = 
\sum_n \vert i,n\rangle\otimes 
U_{\Omega}\overline{\vert i,n \rangle}$ is obtained 
by tracing over the  modules labeled by the primary fields 
$\vert i\rangle = \vert i,0\rangle$. The operator $U_{\Omega = A,B}$ 
selects so-called A- or B-type states by enforcing
charge relations between the `left' and `right' sectors.
Acceptable string states must satisfy Cardy's modular consistency 
conditions, in this case on the annulus.  
For A-type states Cardy's solutions
$\vert\vert i\rangle\rangle_{A} = \sum_j
S_{ij}/(S_{0j})^{1/2}\vert j\rangle\rangle_{A}$ apply. 
They are built from the $S$-matrices, which for minimal models at height $h$ 
(whose primary fields are labelled $\vert i \rangle = \vert l,m,s\rangle$), 
are simple trigonometric functions, whose $SU(2)_k$ part is just 
$S_{l,l'} = \sqrt{2/h}\sin(\pi(l+1)(l'+1)/h)$.
The construction can be adapted to B-type states by restricting
the sum to run over labels appropriate for this case, and extended 
to Gepner models by tensoring such minimal model states \doubref\RS\FS.
Consequently, every detail about these classes of rational boundary 
states in Gepner models is known.  A- and B-type boundary states 
$\vert\vert LMS\rangle\rangle_{\Omega}$ are labeled by vectors 
$L,M$ of $SU(2)$ quantum numbers from the minimal models in the 
tensor product, as well as a vector $S$ which distinguishes
the NS- and R-fields.  We shall return to the classification of 
boundary states in Gepner models elsewhere, since this will be 
important in establishing a complete correspondence between
non-perturbative string theory and the geometry of $D$-branes in CY spaces.

The microscopic version of the intersection numbers in 
classical geometry should be `overlap integrals' (open string
amplitudes) between boundary states $\vert\vert LMS\rangle\rangle_{\Omega}$
and $\vert\vert \tilde L\tilde M\tilde S\rangle\rangle_{\Omega}$.  
As explained in ref. \BDLR\  these can be computed by evaluating 
the Witten index in the Ramond sector.  
The result depends on the parity of $n + r$. When $n + r$ is even 
the intersection matrix for B-type boundary states is:
$$I^{Be} =\  _B\langle\langle\tilde L\tilde M\tilde S \vert\vert I^{e} 
\vert\vert LMS\rangle\rangle_B = C_e 
 \prod_{i=1}^r \sum_{l_i=1}^{2h_i}
 \delta^{(K)}_{l/2,(\tilde M - M)/2}
 \prod_{j=1}^r N_{L_j\tilde L_j}^{l_j},
\eqn\IBe
$$
where  $C_e(S-\tilde S)$ is a known `constant', 
$M = \sum_j w_j M_j = 1,\dots, 2K$, 
$l = \sum_j w_j l_j$, and the $SU(2)_k$ fusion coefficients are 
given by the Verlinde formula $N_{L\tilde L}^{l} = \sum_{l'}
S_{l',L}S_{l',l}S^*_{\tilde L,l'}/S_{l',0}$. The clever trick \BDLR\ is
to extend the definition of the fusion  coefficients in a formal but
natural way (compatible with the Verlinde formula) so that they become
periodic over the extended range of labels.   By virtue of the
properties of the S-matrices and the Verlinde formula  we have
$$
N_{L,\tilde L}^l =  \cases {
\phantom{-}1 & for\quad 
$l = \vert L-\tilde L\vert,\vert L-\tilde L\vert+2,\dots,(L+\tilde L)$\cr
-1 & for\quad 
$l =-\vert L-\tilde L\vert-2,-\vert L-\tilde L\vert-4,\dots,
-(L+\tilde L+2)$\cr
\phantom{-}0 & otherwise \cr}
\eqn\fuse
$$
Eq.\IBe\ coincides with the formula derived in ref. \BDLR.  
When $n + r$ is odd, we find a similar expression
$$
\eqalign{I^{Bo} &=\ 
_B\langle\langle \tilde L\tilde M\tilde S\vert\vert I^{o}  
\vert\vert LMS\rangle\rangle_B \cr
&= C_o \prod_{i=1}^r \sum_{l_i=1}^{2h_i} (-1)^{(M-\tilde M + l)/K }
 \d^{(K/2)}_{l/2,(\tilde M - M)/2}
 \prod_{j=1}^r N_{L_j \tilde L_j}^{l_j},\cr}
\eqn\IBo
$$
but with half the period in the $\delta$-function and an additional sign
factor.

Gepner models enjoy a large group of discrete symmetries which play a
fundamental role in both the structure of the CFT and its geometrical
limits. Since the intersection matrices are computed at the Gepner
points they  inherit many of these symmetries. This can be exploited to
give a much  simpler representation of the intersection matrix in terms
of  two (highly reducible) representations of the abelian group
$\ZZ_K$. Define the fundamental $K$-dimensional `shift' matrices:
$$
\gamma_{\pm}(K)  =  \left( 
\matrix{ 0 & 1 & 0 & \cdots & 0 & 0 \cr
         0 & 0 & 1 & \cdots & 0 & 0 \cr
         \vdots & \vdots & \vdots & \ddots & \vdots & \vdots \cr
         0 & 0 & 0 & \cdots & 0 & 1  \cr
     \pm 1 & 0 & 0 & \cdots & 0 & 0} \right)_{K\times K} , 
\eqn\shift
$$
and note that $\gamma_+(K)$ has period $K$ while 
$\gamma_-(K)$ has period $2 K$, because $\gamma_{\pm}(K)^K = \pm 1$.
For fixed values of $L,\tilde L$ the intersection form can be regarded 
as a matrix with rows and columns labelled by $\tilde M$ and $M$.
When $n + r$ is even we find that $I^B$ is most simply expressed as a 
polynomial in $g =\gamma_+(2K)^2$, while when $n + r$ is odd we find 
that $I^B$ is most simply expressed as a polynomial in $g = \gamma_-(K)^2$.
Introducing the notation $g_j =g^{w_j}$ 
and the fundamental step-operators from ref. \DDCR,
$$
t_{L_j} = t^t_{L_j} = \sum_{l= - L_j/2}^{L_j/2} g_j^l\ ,
\eqn\step
$$
it is possible to prove \DDCR\ that the nasty looking expression for 
$I^{Be}$ displayed above reduces to:
$$
I^B_{L\tilde L} = \prod_j n_{L_j \tilde L_j}
\qquad {\rm with}\qquad 
n_{L_j \tilde L_j} = 
t_{L_j}t_{\tilde L_j} (1 - g_j^{-1}).
\eqn\mangos
$$
We find exactly the same form also for the odd case, even if the
representation involving $\gamma_-$ is used. The particular
representation of the intersection form chosen here emerges naturally
from the BCFT, through the `doubling trick' for fusion coefficients
described above, but it is highly redundant. There are many non-trivial
relations between the set of boundary states $\vert\vert
LMS\rangle\rangle_B$, and this is reflected in the redundancy encoded
in $I^B$.  Similarly, the boundary state itself is redundantly labelled
by $M$, or equivalently the $U(1)$ charges $q_j = M_j/h$ of the
minimal model states from which it is built.   For the spinless sector
we write:  $\vert\vert q\rangle\rangle_B = \vert\vert
L=0;q\rangle\rangle_B$, while higher spin states (not all independent)
are again given by the step-operators: $\vert\vert L;q\rangle\rangle_B
= \prod_j t_{L_j}\vert\vert q\rangle\rangle_B$.

The number of moduli of these boundary states are also given by the 
(extended) fusion coefficients \BDLR
$$
\mu^B_{L} = 
{1\over 2}\prod_j \vert n_{L_j L_j}(\vert g\vert)\vert - \nu\ ,
\eqn\cftmodul
$$ 
where subtracting $\nu$ compensates for the counting of vacuum states.
If $n+r$ is odd, $\nu = 2^\ell$, where $\ell$ is the number of
$L_j$'s equal to  $k_j/2$ \DDCR. However, if $n+r$ is even, we find that
$\nu = 2^{\ell-1}$ when $\ell>0$, and $\nu = 1$ if $\ell = 0$.  Hence
if all $k_j$ are odd, $\nu$ is always equal to one.

Note that from a CFT point of view we are free to extend the tensor
product with trivial ($k_j=0$) factors, if we so desire.  Thus,
although  $\gamma_{-}(K)^2$ provides the simplest and most obvious
representation  when $n+r$ is odd, we can also work in the $n+r+1$ even
channel where  computing the intersection form and number of moduli
requires appending  a factor of $P_{\pm}(g) = (1 \pm g^{K/2})/2$,
appropriate for a minimal  model with no central charge. Since
$P_{+}(g)$ and $P_{-}(g)$ are (orthogonal) projection operators the 
number of such factors is immaterial. This appears to be the reason why the
calculations reported in \DDCR\  are correct.  As already explained, it
is the even channel which  offers the most direct down-link to
geometry.  It is a useful consistency check to compare results from
computations in both channels, and we indeed found agreement.

We now want to transport the CFT data, in particular the charges 
of the B-type boundary states, to large radius points on the boundary 
of moduli space, where they can be compared with the Ramond charges of 
$D$-branes wrapping cycles of large CY manifolds.  This can not be done 
without specifying some geometric data which select a CY space.

\chapter{Bundle data and brane charges}

We summarize the relevant $D$-brane geometry for the $K3$-fibrations
under examination here. \Xtwo\ has been analyzed in great detail using 
mirror symmetry in ref. \COFKM, who also discussed \Xone\ in somewhat less 
detail; further explicit properties of the latter were worked out 
in ref. \KKLMV.  For each of the two compactifications
the $N=2$ special geometry is encoded in a prepotential 
whose classical part can be written in the form:
$$
\cF(t_1,t_2)\ =-\Coeff1{3!} c_{111}\to^3-\Coeff12 c_{112}\to^2\tz+
\Coeff1{24}(b_1\to+b_2\tz)+{\rm const.},
\eqn\prepot
$$
where $t_1$ and $t_2$ parametrize the complexified K\"ahler cone:
$K=t_1J_1+t_2J_2$. Here $J_1$ and $J_2$ are the K\"ahler classes of the
$K3$-fiber and $\IP_1$ base, respectively.\foot{More precisely,
$J_1=2L+E$ and $J_2=L$, where $E$ is an exceptional divisor coming from
blowing up a curve of singularities, and the linear system $|L|$ is a
pencil of $K3$'s.} The coefficients are given in terms of the following
topological intersections:
$$
c_{ijk}\ =\ \int J_i\wedge J_j\wedge J_k\ ,\qquad
b_i\ =\ \int c_2(X)\wedge J_i\ , \ i=1,2.
\eqn\intersections
$$
For the threefolds at hand, we have \COFKM:
$$
\eqalign{
\xone:\ \qquad &c_{111}=4,\ c_{112}=2,\ b_1=52,\ b_2=24 \cr
\xtwo:\ \qquad &c_{111}=8,\ c_{112}=4,\ b_1=56,\ b_2=24 \cr}
\eqn\topdata
$$
From the prepotential one derives the symplectic period vector
$\Pi\equiv(\cF_0\equiv2\cF-t_i\del_i\cF,\cF_1,\cF_2,1,t_1,t_2)^t$,
which for \Xone\ reads:
$$
\Pi\ =\ 
\pmatrix{ {\frac{13\,{t_1}}{6}} + {\frac{2\,
{{{t_1}}^3}}{3}} + {t_2} + 
   {{{t_1}}^2}\,{t_2} \cr
 {\frac{13}{6}} - 2 {t_1}^2- 2\,{t_1}\,{t_2} \cr 1 - 
   {{{t_1}}^2} \cr 1 \cr {t_1} \cr {t_2} \cr  }
\eqn\periods
$$
The entries (from the top down) correspond to integrations over  
$6-,4-,4-,0-,2-$ and $2-$cycles, respectively.

We now want to obtain an explicit map between the topological
invariants of $K$-theory (characteristic classes of the Chan-Paton
sheaf $V$), and brane charges.  The latter are defined as the
coefficients of the $N=2$ central charge given by
$$
Z(n_i)\ =\ \nn6\, \cF_0+\n41\, \cF_1+\n42 \,
\cF_2+\nn0 +\n21 t_1+\n22 t_2\ .
\eqn\nCentral
$$
Following ref. \DDCR, the topological invariants can be obtained by comparing
this to the central charge of a $D$-brane configuration
$$
Z(Q(V))\ =\ -\int e^{-K}\!\wedge Q\ =\ 
-\int e^{-K}\!\wedge \tr(e^F)\wedge \sqrt{\hat A(X)}\ ,
\eqn\QCentral
$$
where $Q=(r,\ c_1(V),\ \chv2+{r\over24} c_2(X),\
\chv3+{1\over24}c_1(V)c_2(X)) \in H^{2*}(X)$ is the Mukai vector of
effective brane charges \doubref\GHM\HM. Specifically, $r$ is the rank
of the bundle and $\chv i$ are its Chern characters. Inserting the
classical periods \periods\ and substituting the values of the
intersections  given in \topdata, equating \nCentral\ and \QCentral\
yields:
$$
\eqalign{
r\     &=\ n_6\ ,\quad
c_1(V)\  =\ \Coeff12 \n41 J_1 +(\n42-\n41) J_2 \cr
\chv2\ &=\  n^1_2\, h + n^2_2\, \ell\ ,\quad
\chv3\  =\ -\nn0-\frac13 \n41-2 \n42\ .}
\eqn\bundledata
$$
where $h=\shalf J_1\wedge J_2$ and $\ell= \shalf J_1\wedge J_1-
J_1\wedge J_2$.

We will be mainly interested in $D$-branes wrapping the $K3$-fiber
(called ``fiber branes'' in ref. \HM). This corresponds to considering only
states with  $\n41=\n22=0$, and accordingly we will set $\nn4\equiv
\n42$ and  $\nn2\equiv\n21$. It follows from ref. \DDCR\ that the
topological invariants of the torsion sheaf $V$ supported on the  
$K3$-fiber can be written in terms of its extension $i_*V$ to the 
threefold as:
$$
Q\ =\ (0,r\,J_2,i_*c_1(V),i_*\chv2+r)\ .
\eqn\restrQ
$$
That the vector of induced charges takes this naive form  is a
consequence of the triviality of the normal bundle, $c_1(K3)=0$, and
$c_2(K3)=24$; otherwise, there would be extra correction terms.
Comparing the central charges as before, we thus obtain a simple
truncation of \bundledata:
$$
r\ =\ \nn4\ ,\quad i_*c_1(V)\ =\ \nn2\  h\ ,\quad 
i_*\chv2\ =\ -\nn0-2\nn4\ .
\eqn\bundledata
$$

We are particularly interested in the dimensions of the moduli spaces
$\cM$ of $D$-brane configurations on the $K3$-fiber. Unlike for
threefolds, for an isolated $K3$   such dimensions (for irreducible
bundles) are completely determined by the Mukai vector~\mukai:
$$
\mu_{K3}(Q)\ \equiv\ {\rm dim}_{c}\cM(Q)\ =\ \langle Q,\,Q\rangle+2\ .
\eqn\dimMQ
$$
To write this in terms of the brane charges $n_i$, we need to know
what the precise form of the inner product in \dimMQ\ is. For this,
note that the intersection form of the $K3$-fiber is easily exhibited by
transforming the periods \periods\ in the $t_2\rightarrow\infty$ limit
to a more suitable basis, given by $\tilde\Pi^t=(1,\to,\to^2,\tz,\tz\to,
\tz\to^2)^t$. The symplectic intersection form on the threefold then
turns into $\Interg\rightarrow\tilde \Interg= \left({{\,0\ \ \
{\!\!1\,}}\atop {{\!-\!1}\ 0}}\right)\otimes \Inters_{K3}$, where
$$
\Inters_{K3} \ = 
\pmatrix{ 0 & 0 & -1 \cr 0 & {\frac{1}{2}} & 0 \cr -1 & 0 & 0 \cr  }\ .
\eqn\OmKt
$$
Accordingly, the Mukai charge vector on $K3$ takes the form
$Q_{K3} = (r,c_1(V),\chv2+r) = (\nn4,\nn2,-\nn0-\nn4)$,
whence:
$$
\mu_{K3}(n_i)\ =\ 2+\Coeff12\nn2^2+2 \nn0\nn4+2\nn4^2\ .
\eqn\dimMn
$$
Analogously, for the threefold \Xtwo\ we find that the intersection
form on the fiber is half of the one given in \OmKt, and
$Q_{K3}=(2r,c_1(V),\chv2+r)=(2\nn4,\nn2,-\nn0-\nn4)$. We therefore
get in this case:
$$
\mu_{K3}(n_i)\ =\ 2+\Coeff14\nn2^2+2 \nn0\nn4+2\nn4^2\ .
\eqn\dimMnei
$$

\chapter{Matching of geometric and BCFT data}

We now relate the BCFT data of Sect. 2 to the geometrical data of the
previous section. This first of all involves an analytic continuation
of the large-radius periods \periods\ to the Gepner point, i.e. 
$Z=\vec n\cdot\vec\Pi= (\vec n^G\, m^{-1})\cdot (m\,\vec\Pi^G)$. 
The continuation matrix $m$ is path dependent and thus only defined up to
$Sp(6,\ZZ)$ monodromy transformations. It is most natural to choose,
similar to the case in refs. \doubref\COFKM\DDCR, a preferred basis such
that the conifold singularity corresponds to a single wrapped
$D6$-brane.

By applying the method for analytic continuation described in ref.  
\COFKM\ to the threefold \Xone, we find after some work that 
for our choice of periods \periods\ the continuation matrix takes the form
$$
m\ =\ \pmatrix{ -1 & 1 & 0 & 0 & 0 & 0 \cr {\frac{3}{2}} & {\frac{3}{2}} & 
    {\frac{1}{2}} & {\frac{1}{2}} & -{\frac{1}{2}} & -{\frac{1}
     {2}} \cr 1 & 0 & 1 & 0 & 0 & 0 \cr 1 & 0 & 0 & 0 & 0 & 0 \cr -{
      \frac{1}{2}} & 0 & {\frac{1}{2}} & 0 & {\frac{1}{2}} & 0 \cr {
     \frac{1}{2}} & {\frac{1}{2}} & -{\frac{1}{2}} & {\frac{1}
    {2}} & -{\frac{1}{2}} & {\frac{1}{2}} \cr  }\ .
\eqn\Mone
$$
The geometric intersection form in the Gepner basis is therefore:
$$
\Inters^G = m^{-1}\, \Interg\, m^{-1,t}\ =\ 
\pmatrix{ 0 & 1 & 0 & -2 & 0 & 1 \cr -1 & 0 & 1 & 0 & -2 & 0 \cr 0 & 
    -1 & 0 & 1 & 0 & -2 \cr 2 & 0 & -1 & 0 & 1 & 0 \cr 0 & 2 & 0 & 
    -1 & 0 & 1 \cr -1 & 0 & 2 & 0 & -1 & 0 \cr  } \ .
\eqn\Igep
$$
On the other hand, for \Xtwo\ we get\foot{Note that we use a basis
different to the one of ref. \COFKM; moreover, we found that there are
some typing errors in their monodromy matrices.}
$$ m\ = \
\pmatrix{ -1 & 1 & 0 & 0 & 0 & 0 \cr {\frac{3}{2}} & {\frac{3}
    {2}} & 0 & 0 & -{\frac{1}{2}} & -{\frac{1}
     {2}} \cr 1 & 0 & 1 & 0 & 0 & 0 \cr 1 & 0 & 0 & 0 & 0 & 0 \cr -{
      \frac{1}{4}} & 0 & {\frac{1}{2}} & 0 & {\frac{1}{4}} & 0 \cr {
     \frac{1}{4}} & {\frac{3}{4}} & -{\frac{1}{2}} & {\frac{1}
    {2}} & -{\frac{1}{4}} & {\frac{1}{4}} \cr  }
\eqn\Mtwo
$$
and an expression for $\Inters^G$ similar to \Igep.

Since there is no way the BCFT can know how we will choose the
continuation path for a given large-radius `endpoint' threefold, the
monodromy  basis selected by the above construction of boundary states
need not be the same as the Gepner basis associated with our choice  of
$m$.  A further change of basis is necessary to link  the
BCFT boundary state charges to the  geometrical charges at  the Gepner
point. The precise form of the required transformation can be determined by
comparing the geometric intersection matrix at the Gepner point \Igep\
with the projection of the $L_i=0$ BCFT intersection matrix \IBe, \IBo\
onto the non-redundant basis.

In all cases studied in refs. \doubref\BDLR\DDCR\ and by us so far, it turns
out that $I^B$ is to be compared with $(1-g)T I^GT^t(1-g)^t$,
where $T$ is the appropriate intertwining matrix. Taking
everything together, the geometric brane configurations  $\vert
n_i\rangle^{\infty}$  are given in terms of the boundary CFT charges by 
$\vert n_i\rangle^{\infty} =  (m^{-1})^{t}\, T^t (1-g)^t\vert\vert
L;q\rangle\rangle_B$, where $n_i$ can be thought of as functions of
the BCFT labels $\{ L;q\}$.

What we find for the two models considered here are lists of charges
that are similar to those of ref.\ \DDCR. Indeed a similar bundle
interpretation can be given for some of them, however in order not to
be repetitious, we refrain from presenting this here. There are also certain
states, like the one with charges $(1,-2,0)$, for which there is no
conventional bundle interpretation, and this is analogous to the
findings of \BDLR.

The major difference of our work and ref.\ \DDCR\ is that we deal  with
$K3$-fibrations rather than elliptic fibrations.  We will thus focus
here on the brane features which are specific to $K3$ fibrations, and
in particular on brane configurations supported on the $K3$ fiber.
Tables 1 and 2 list all brane states we have found within this
construction that have $n_6=\n41=\n22=0$, together with some of their
essential characteristics.

\goodbreak
\def\ft[#1,#2,#3]{(#1,#2,#3)}
\def\fl[#1,#2,#3,#4]{[#1,#2,#3,#4]}
\vskip 1.truecm
\begintable
$L_i$  &  $\nu$  | & $\!\!\!\!\big(r,\chv2,c_1(V)\big)\!\!\!\!$ & |  $\mu^B_L$ 
 &  $\Delta$  \crthick
\fl[1,0,0,0] & 1 | \ft[1,0,0]  & \ft[1,1,-2]  & \ft[2,-1,-2] | 1 & 1\cr 
\fl[3,0,0,0] & 1 | \ft[1,-2,0] & \ft[1,-1,-2] & \ft[0,-1,2]  | 5 & 1\cr 
\fl[3,0,1,0] & 1 | \ft[1,0,-4] & \ft[1,-3,2]  & \ft[2,-3,-2] | 9 & 1\cr
\fl[3,0,1,1] & 1 | \ft[3,-6,0] & \ft[3,-3,-6] & \ft[0,-3,6]  | 21 & 1\crthick 
\fl[5,0,0,0] & 2 | \ft[2,-2,0] & \ft[2,0,-4]  & \ft[0,-2,0]  | 6 & 4\cr
\fl[5,0,1,0] & 2 | \ft[2,-4,0] & \ft[2,-2,-4] & \ft[0,-2,4]  | 14 & 4\cr 
\fl[5,0,1,1] & 2 | \ft[2,0,-8] & \ft[2,-6,4]  & \ft[4,-6,-4] | 30 & 4\crthick  
\fl[5,0,2,0] & 4 | \ft[4,-4,-4] & \ft[0,-4,4]  & \ft[0,0,4]  | 20 & 10\cr
\fl[5,0,2,1] & 4 | \ft[4,-8,0]  & \ft[4,-4,-8] & \ft[0,-4,8] | 44 & 10\crthick 
\fl[5,0,2,2] & 8 | \ft[4,0,-12] & \ft[4,-8,-4] & \ft[4,-8,4] | 64 & 22
\endtable
\noindent{\bf Table 1:}
{\sl
CFT and bundle data of states localized on the $K3$-fiber of  \Xone. On
the left we list the CFT labels $L_i$ and the number of vacuum states
$\nu$, as defined in the text after eq.\ \cftmodul\ (note that in some
cases there are two sets of such labels that lead to the same states,
and we show only one of them). In the middle we have three sets of
charges,  which have the same properties because they
belong to the same $\ZZ_{12}$ orbit. On the right, $\mu^B_L$ is the
number of BCFT moduli in the threefold as given by eq.\ \cftmodul.
Finally, $\Delta\equiv\mu^B_L - \mu_{K3}$ is the excess number of
moduli of the boundary state as compared to the number of
$K3$ bundle moduli $\mu_{K3}$ given by eq.\dimMn.
} \goodbreak

\def\ft[#1,#2,#3]{(#1,#2,#3)}
\def\fl[#1,#2,#3,#4,#5]{[#1,#2,#3,#4,#5]}
\begintable
$L_i$  &  $\nu$  | \multispan{4} $\big(r,\chv2,c_1(V)\big)$ |  $\mu^B_L$ 
 &  $\Delta$  \crthick
\fl[1,0,0,0,0] & 1 | \ft[1,0,0] & \ft[3,0,-8] & \ft[1,2,-4] & 
\ft[3,-2,-4] | 1 & 1\cr 
\fl[3,0,0,0,0] & 1 |  &  \ft[2,-2,-4] & \ft[0,-2,4]  &  | 7 & 1\crthick
\fl[3,0,1,0,0] & 2 |  &  \ft[2,-4,0]  & \ft[2,0,-8]  &  | 14 & 4\crthick 
\fl[3,0,1,1,0] & 4 |  &  \ft[4,-4,-8] & \ft[0,-4,8]  &  | 28 & 10\crthick
\fl[3,0,1,1,1] & 8 |  &  \ft[4,-8,0]  & \ft[4,0,-16] &  | 56 & 22
\endtable
\noindent{\bf Table 2:}
{\sl
Analogous data as in Table 1 for $D$-branes wrapped on the $K3$-fiber of \Xtwo.
} 

\goodbreak

\chapter{Discussion}

The data in Tables 1 and 2 provide support for the conjectured
equivalence of boundary states and $D$-branes. We confine ourselves here to 
the following remarks.

\ni {\bf i)} The entries of each row belong to the same $\ZZ_K$ orbit
(which also contains other states not localized only on the $K3$ fiber
that we don't show). They thus have the same properties with regard to
the moduli space, even though the brane charges can be very different.
All charges satisfy the BPS condition \HM\ $\langle
Q_{K3},Q_{K3}\rangle\geq-2$, with $r> 0$ or $r=0,c_1>0$ or $r=c_1=0$,
$\chv2\equiv \shalf {c_1}^2-c_2<0$. \footnote{1\ }{Note that  $\chv2<0$
corresponds to an anti-selfdual gauge connection,  and positive brane
charges, cf.\ eq.\ \bundledata.}

\ni {\bf ii)} Consider the charge vectors of the form
$Q_{K3}=(r,-c_2,0)$, which can be interpreted as $SU(r)$ bundles with
$c_1=0$.  For such bundles $c_2$ must obey $c_2\geq 2r$. We indeed
find, as a first consistency test, that for all such charges, $c_2$ is
equal to $2r$, except for $r=2$ where we find in addition also $c_2=r$.
However, this can be given an interpretation as an $SO(3)$ bundle
instead of $SU(2)$, so all is well.

\ni {\bf iii)}
More interestingly, note that for boundary states with one vacuum
($\nu=1$ in \cftmodul), the number of CFT moduli is one larger
than the number of classical $K3$ moduli: $\Delta\equiv
\mu^B_L-\mu_{K3}=1$. This is indeed exactly as expected from geometrical
considerations: the brane system should gain a modulus
from the one-dimensional base space ($\IP_1$ here) when it is 
embedded as a fiber in a threefold.

\ni {\bf iv)} On the other hand, if the number of CFT vacua gets
larger, $\nu>1$, there is a growing discrepancy between the numbers of
threefold BCFT moduli and classical $K3$ moduli. However, as suggested
in \DDCR, such boundary states may describe not single branes but
collections of several branes, and this would correspond to reducible
bundles or sheaves. For such bundles the geometry of the moduli space
is more intricate than for irreducible ones, as the relative positions
of the configurations correspond to Coulomb branches of additional
moduli \BS.


By studying the tables one finds consistency between the charge
vectors and the values of $\nu$. That is, configurations with $\nu>1$
can be decomposed into configurations with smaller $\nu_i$, in a way
such that $\nu_i$ and the charges add up correctly.\foot{Details will
be discussed elsewhere.} For example, the configurations with charges
$(2,-2k,0)$, $k=1,2$ have $\nu=2$, and indeed can be decomposed \BS\
into two branes with charges $(1,\shalf {c_1}^2(L),\pm c_1(L))$. These
correspond to reducible bundles of the form $L\oplus L^{-1}$, where $L$
are line bundles with ${c_1}^2(L)=-2k$. The Higgs branches of these
$U(1)\times U(1)$ theories happen to coincide \BS\ with the moduli
spaces of $SU(2)$ bundles with charges $(2,-2k,0)$.  This
illustrates that the charge label alone does not completely specify the
configuration, rather it is $\nu$ that tells that we have a reducible
$U(1)\times U(1)$ bundle rather than an $SU(2)$ bundle (however, these
configurations are continuously connected, so that a distinction is not
fundamentally important.) In contrast, the configuration $(3,-6,0)$ in
table~1 is irreducible because $\nu=1$. Therefore it corresponds to an
$SU(3)$ bundle with no Coulomb branch.

Some interesting observations can also be made by comparing both
tables: for a given number $\nu$ of CFT vacuum states, we
find that the number of excess moduli $\Delta$ is always the same. For
example, the state with charges $(4,-4,-8)$ has 44 CFT moduli on \Xone,
which is 10 more than what the Mukai formula gives for the $K3$ alone.
On the other hand, it has 28 moduli\foot{Note that the two $K3$'s have
different intersection forms, so the Mukai formulas \dimMn\ and
\dimMnei\ yield different numbers of moduli.} on \Xtwo, which is again
10 more than one would have for the corresponding $K3$.

The point is that while the absolute numbers of moduli may differ, the
excess number of moduli coming from the fibration  is universal and
given by\foot{This is a general property of the BCFT, as will be shown
elsewhere.} $\Delta=3\nu-2$. This can be given the following
interpretation. From index theorems we know that
$\mu_{K3}=N_m-2(N_g-1)$, where $N_m=2r\!\int\! c_2$ is the number of
matter fields in six dimensions and $N_g$ is the total number of gauge
fields. For an irreducible bundle we have $N_g={\rm dim}G+1$,
where the shift by one is due to the $U(1)$ factor that corresponds to
the overall center-of-mass modulus of the brane configuration \BDLR.
For a reducible bundle, which corresponds to a collection of branes, we
have several center-of-mass moduli and thus $N_g={\rm dim}G+N_{U(1)}$,
where $N_{U(1)}$ is equal to the number of irreducible components.
Hence we can identify $N_{U(1)}=\nu$ and express the excess number of
moduli as:
$$
\eqalign{
\Delta\ \equiv\ \mu^B_L-\mu_{K3}\ &=\ \mu^B_L-(N_m-2(N_g-1))
\cr
&=\ \Big(\mu^B_L-(N_m-2{\rm dim}G)\Big)+ 2(\nu-1)\cr
&=:\ \tilde \Delta +  2(\nu-1)\ .
}\eqn\tildedelta
$$
The second term may be thought of as a correction to the Mukai formula
$\mu_{K3}=2r\!\int\! c_2-2{\rm dim}G$,  taking into account that we
have a reducible bundle if $\nu>1$.  A comparison with the CFT result
$\Delta=3\nu-2$ then yields
$$
\tilde \Delta\ =\ 
\sum_{{\rm irreducible \atop components}}^\nu \tilde\Delta_{(i)}
\ =\ \nu\ ,
\eqn\nice
$$
which means that each irreducible component of the $K3$ bundle or sheaf
gains one extra modulus upon fibration -- which is precisely  as expected !
This is because each such component corresponds to an independent brane
configuration that can sit anywhere over the $\IP^1$ base.

Concluding, we note that there is a high degree of
consistency and universality in our results. We therefore believe that 
these provide a successful test of the BCFT approach to $D$-branes.

Our findings suggest in particular that the moduli space of the $K3$ remains
protected even when embedded in a threefold with less supersymmetries,
and its dimension is still given by the Mukai formula \dimMQ,
if we subtract the degrees of freedom associated with the fibration over
the $\IP_1$ base. While this robustness may not be too surprising in an
adiabatic, large base limit of the threefold, it is more impressive
here as we look into a region of the threefold moduli space (the Gepner
point) where the classical geometry is maximally distorted by quantum
corrections.  It appears that nothing serious happens to the branes
supported on the $K3$-fiber at large radius when we transport them all
the way to the  Gepner point of the CY moduli space.  One may
argue that there is an $N=4$ subsector in the $N=2$ BCFT that protects
these states from instanton corrections also in the non-geometric regime.

\goodbreak 
\ack
\nobreak
We would like to thank  J\"urg Fr\"ohlich, J\"urgen Fuchs, 
Michael Gutperle, Graham Ross, Christoph Schweigert, and especially 
Peter Mayr for useful discussions, and Emanuel Diaconescu
for illuminating correspondence.

\bigskip
\goodbreak
\refout
\end